\documentclass[journal=nalefd,manuscript=letter,layout=traditionnal]{achemso}

\usepackage[version=3]{mhchem} 

\author{B. Voisin}
\affiliation{Univ. Grenoble Alpes, INAC-SPSMS, F-38000 Grenoble, France}
\alsoaffiliation{CEA, INAC-SPSMS, F-38000 Grenoble, France}
\author{R. Maurand}
\email{romain.maurand@cea.fr}
\affiliation{Univ. Grenoble Alpes, INAC-SPSMS, F-38000 Grenoble, France}
\alsoaffiliation{CEA, INAC-SPSMS, F-38000 Grenoble, France}
\author{S. Barraud}
\author{M. Vinet}
\affiliation{CEA, LETI, MINATEC Campus, 17 rue des Martyrs, 38054 Grenoble, France}
\author{X. Jehl}
\affiliation{Univ. Grenoble Alpes, INAC-SPSMS, F-38000 Grenoble, France}
\alsoaffiliation{CEA, INAC-SPSMS, F-38000 Grenoble, France}
\author{M.Sanquer}
\affiliation{Univ. Grenoble Alpes, INAC-SPSMS, F-38000 Grenoble, France}
\alsoaffiliation{CEA, INAC-SPSMS, F-38000 Grenoble, France}
\author{J.Renard}
\affiliation{Univ. Grenoble Alpes, INAC-SPSMS, F-38000 Grenoble, France}
\alsoaffiliation{CEA, INAC-SPSMS, F-38000 Grenoble, France}
\author{S. De Franceschi}
\affiliation{Univ. Grenoble Alpes, INAC-SPSMS, F-38000 Grenoble, France}
\alsoaffiliation{CEA, INAC-SPSMS, F-38000 Grenoble, France}

\title[An \textsf{achemso} demo]
  {Electrical control of g-factors in a few-hole silicon nanowire MOSFET}
\abbreviations{MOSFET, SOI, SO-interaction}
\keywords{Quantum dot, Hole transport, Lande g-factor, Silicon, MOSFET}

\begin{document}


\begin{abstract}
Hole spins in silicon represent a promising yet barely explored direction for solid-state quantum computation, possibly combining long spin coherence, resulting from a reduced hyperfine interaction, and fast electrically driven qubit manipulation. 
 
Here we show that a silicon-nanowire field-effect transistor based on state-of-the-art silicon-on-insulator technology can be operated as a few-hole quantum dot. A detailed magnetotransport study of the first accessible hole reveals a g-factor with unexpectedly strong anisotropy and gate dependence. We infer that these two characteristics could enable an electrically-driven g-tensor-modulation spin resonance with Rabi frequencies exceeding several hundred MHz.
\end{abstract}


Localized spins in semiconductors provide a natural way to encode quantum information in a solid-state environment~\cite{Loss1998,Kane1998}. This opportunity has been extensively investigated using the spin of electrons confined to quantum dots in GaAs/AlGaAs high-mobility 
heterostructures~\cite{Koppens2005,Johnson2005,Petta2005}. In this system, and III-V materials in general, the electron spin coherence is rather short though. The presence of a large hyperfine coupling with the nuclear spins of the host crystal elements leads to time averaged coherence times, $T_2^*$, of only a few tens of ns. This limitation can be overcome by considering spin-qubit implementations in group-IV semiconductors (Si, Ge, C), where the concentration of isotopes carrying non-zero nuclear spins is naturally much smaller. As a matter of fact, the concentration of group-IV elements with non-zero nuclear spin can be even lowered by means of isotopic purification techniques~\cite{Itoh2003}.  
An outstanding  $T_2^*$ of 120 $\mu$s, i.e. four orders of magnitude longer than in III-V materials, was obtained in a spin-qubit device made out of isotopically purified silicon, with over 99.9 \% of nuclear-spin-free $^{28}$Si, the most abundant silicon isotope~\cite{vel14}.

Most of the experimental work on silicon-based spin qubits has so-far focused on electron spins, either bound to a donor atom~\cite{pla12} or confined to a gate-defined quantum dot~\cite{vel14,kaw14,Hao2014}.  On the other hand, hole spin qubits have been lagging behind, although they may potentially present some advantages. The first advantage comes from the p-wave symmetry of the hole Bloch functions. Having a node at the atomic site suppresses the direct-contact coupling to the nuclear magnetic moments, which results in a weaker hyperfine interaction~\cite{fis08}. Optical experiments performed on hole spins in self-assembled InAs quantum dots have confirmed this expectation revealing a hole spin coherence time an order of magnitude longer than in the electron case~\cite{Brunner2009}.  The second advantage has to do with the existence in the valence band of a sizable spin-orbit (SO) interaction.    
By coupling spin and orbital degrees of freedom, the SO interaction offers the possibility of fast all-electrical spin manipulation as opposed to magnetic-field driven electron spin resonance.  
The SO mediated approach only requires a microwave modulation on a gate electrode and it can in principle yield very large Rabi frequencies. On the other hand, the magnetic-field driven ESR performed using oscillatory magnetic requires an additional manipulation circuit. This approach is limited by the need to minimize local Joule heating, which translates into rather limited Rabi frequencies, typically below 10 MHz. A third approach enabling all-electrical yet magnetically-driven ESR was successfully demonstrated~\cite{Tok2006,Pio2008}. It uses a micro-magnet to locally create an inhomogeneous magnetic field in which the qubit electron is "shaken" by oscillating a gate voltage. This approach has so far enabled Rabi frequencies as high as $100$ MHz but a recent theoretical work showed that the magnetic field gradient induces extra dephasing~\cite{Beau13}. Finally a last electrical manipulation have been recently demonstrated on a P donor in silicon using Stark shift for coherent rotation~\cite{Laucht2015}.

The precise SO-related mechanism underlying spin manipulation depends on the nature of the SO interaction. In the case of Rashba and Dresselhaus SO interaction, spin rotation is achieved via the so-called mechanism of electric dipole spin resonance (EDSR)~\cite{ now07,bul07},  experimentally demonstrated for electron qubits in GaAs ~\cite{now07}, InAs ~\cite{Nadj-Perge2010}, InSb ~\cite{VandenBerg2013} and carbon nanotube~\cite{Laird2013} quantum dots. The SO interaction can as well result in gate-dependent and anisotropic g-factors. In this case, another possibility arises : spin manipulation via electrically driven g-tensor modulation resonance (g-TMR), as demonstrated for electrons spins  in InGaAs-based heterostructures~\cite{kat03}. 
This latter approach could as well be applied to hole-type quantum dots. A few examples of gate defined hole quantum dots were reported in the recent years~\cite{Dotsch2001,Grbic2005,hu07,Klochan2011,hu12,Prib13} and, to the best of our knowledge,  the only experimental demonstration of a SO-mediated hole-spin resonance was reported for InSb nanowire quantum dots~\cite{Prib13}. The implementation of hole spin qubits in silicon-based quantum dots would be particularly interesting owing to the lower hyperfine coupling and the compatibility with mainstream silicon technology. 

In this letter we show that state-of-the-art metal-oxide field-effect transistors (MOSFETs) can be operated as p-type quantum dots in the few-hole regime. So far, the few-hole regime in quantum dots had only been reached in bottom-up nanowires, based on either InSb~\cite{Prib13} or silicon~\cite{Zhong2005,zwa09}, while the operation of top-down nanostructures was limited to large numbers of holes ~\cite{Dotsch2001,Grbic2005,Spruijtenburg2013, Li2013}.  Using magneto-transport measurements at low temperature, we study the Zeeman splitting of the first hole level accessible evidencing a spin doublet with predominant heavy-hole character. The anisotropy and gate-dependence of the corresponding hole $g$-factor reveal that the studied system meets all the requirements to be used as a spin qubit electrically manipulated by g-TMR. Simulations predict Rabi frequencies exceeding several hundred MHz, more than two orders of magnitudes higher than those of recent works on electron spin qubits~\cite{vel14, kaw14, muh14}.

\begin{figure} 
\includegraphics[width=17cm]{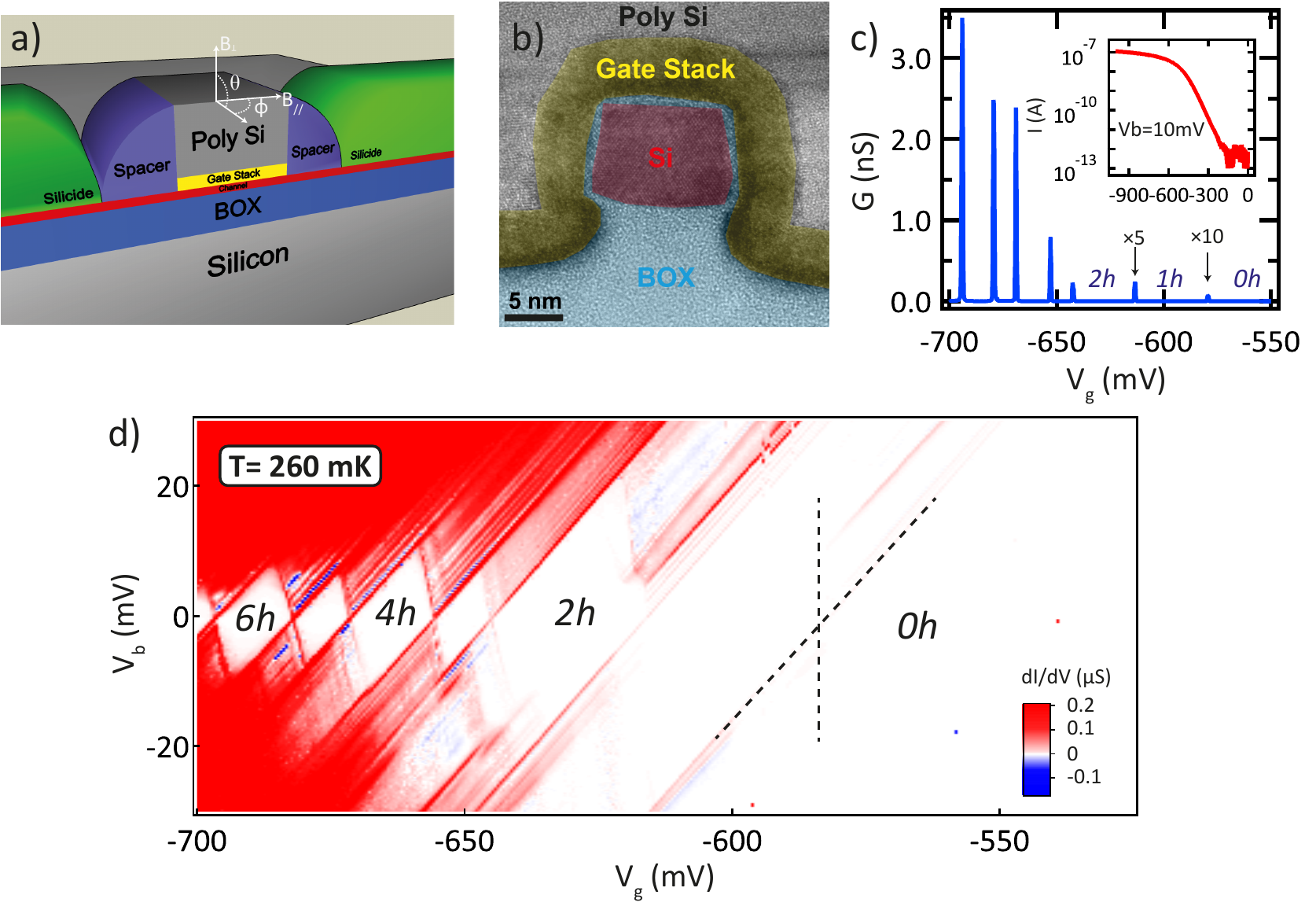} \caption{a) Simplified schematics of a CMOS nanowire transistor. A quantum dot is formed below the gate inside a thin silicon on insolator (SOI) channel. The dot is isolated from the silicided source and drain by the presence of spacers. b) False colored transmission electron microscope (TEM) micrograph of a nanowire transistor section, the silicon channel (in red) at the top of a SOI wafer with a $145$nm buried oxide (BOX) layer (blue) is surrounded by a gate stack of SiO$_2$ and high-$k$ dielectric (in yellow). The gate (in grey) is realized in poly-Silicon. c) Measured differential conductance taken at zero bias as a function of gate voltage ($V_{g}$) at a temperature of $T=4$ K showing Coulomb oscillations. (x5 and x10 indicate that the corresponding current peaks have been multiplied by respectively 5 and 10 times for visibility). Inset shows the transistor characteristic, current versus $V_{g}$ at room temperature at $V_{b}=10$ mV. The sub-threshold slope close to $60$mV per decade indicates an excellent electrostatic control. d) Stability diagram dI/dV vs. ($V_b$, $V_g$) performed at $260$ mK showing the addition of the first holes in the quantum dot. The first resonance appears at the threshold voltage around $V_g=-580$ mV and no feature is observed below.} 
\label{fig1} 
\end{figure}

The measured transistors were fabricated on a complementary metal-oxide semiconductor (CMOS) platform starting from 300-mm-diameter undoped silicon-on-insulator (SOI) wafers ($(100)$ surface orientation). Fabrication involves several steps of optical lithography, metal/dielectric deposition, and etching. The use of an industrial-level CMOS platform presents crucial advantages: an excellent electrostatic control of the undoped channel is achieved despite ultimate characteristic dimensions, with a channel length of about 25 nm oriented along the [110] direction, width and thickness of about 10 nm (see fig.~\ref{fig1}a),b)); the devices benefit from low-resistance source and drain leads fabricated through a boron implantation process followed by silicidation. These two features enable the achievement of the few-hole regime reported here. The high control and reproducibility of CMOS technology is also essential for qubit integration towards large-scale quantum computing architectures.

Devices were studied in a $^3$He measurement setup with a base temperature of 260 mK. A 9-T superconducting magnet and a mechanical sample rotator were used to investigate  the hole g-factors and their angular dependences with an estimated error of $\pm5^\circ$. Electrical measurements were performed using both direct current (DC) and low-frequency lock-in techniques. In our setup, noise attenuation along the DC lines is carried out at room temperature, using pi-filters, and at low temperature, using low-pass RC filters ($\sim 10$kHz cut-off frequency) thermally anchored to the  $^3$He pot. This enables electronic temperatures as low as the cryostat base temperature. Current is amplified at room temperature by means of a battery powered current-voltage converter before DC or lock-in detection.

At $4$K the linear conductance of the device presents peaks above the threshold voltage as shown in fig.~\ref{fig1}(c). This is the signature that charge carriers are accumulated in a single island formed under the gate. The localized holes are weakly coupled to the source and drain leads. Confinement along the wire direction owes to the presence of spacers separating the gated central part of the undoped channel from the heavily doped source and drain leads (see fig.~\ref{fig1}(a),(b)). The stability diagram of the device, i.e. a two-dimensional map of the differential conductance $dI/dV$ as a function of the gate voltage $V_g$ and of the source-drain bias voltage $V_b$, is shown in fig.~\ref{fig1}(d). It shows clear Coulomb-blockade regions, signature of single-hole transport through the quantum dot. Note here that two corner dots~\cite{Voisin2014} are not expected due to the narrow width (W=10nm) of the device. The first diamond is positioned just above the room temperature threshold voltage (see inset of fig.~\ref{fig1}(c)), at which charge carriers start to populate the valence band in the channel. No extra feature appears below this first resonance up to 50 mV in bias (see supplementary material S1) which means that the last hole in the channel has been most probably observed~\cite{zwa09}. Note however that an irrefutable proof would be given by a charge detection method. Details about electronic transport in the regime of a higher number of confined holes are presented in the supplementary material S2.

\begin{figure} 
\includegraphics{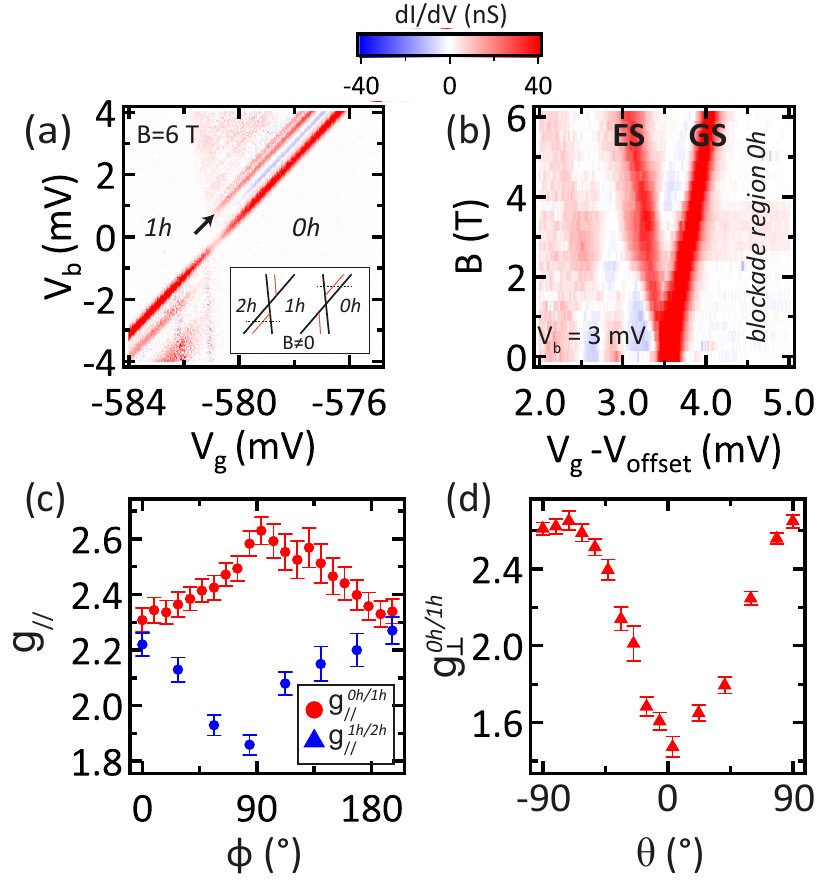} \caption{a) 2D-color plot of the differential conductance $dI/dV$ as a function of bias and gate voltages at a magnetic field of $B=6$T at the zero to one hole transition. The black arrow indicates the conductance resonance through the Zeeman split excited state. The inset sketches a stability diagram around the one hole charge state at finite magnetic field. Excited states resonances are plotted in red, while the dashed lines indicate schematically where the magnetospectroscopies have been recorded. b) 2D-color plot of the differential conductance $dI/dV$ as a function of gate voltage and magnetic field recorded at a fixed bias voltage of $V_{b}=3$ mV. The gate voltage scale is $V_{g}-V_{offset}$ with $V_{offset}=-581$ mV. The linear-in-field splitting of the conductance peaks is a consequence of the Zeeman effect. c) Magnetic field orientation dependence of the $g$ factor for the zero to one hole (in red indicated as $g_{//}^{0h/1h}$ and one hole to two holes (in blue indicated as $g_{//}^{1h/2h}$) transitions. An angle of $\phi=0^{\circ}$ corresponds to a magnetic field applied along the channel of the transistor. d) Same as c) but for the out-of-plane magnetic field measured at
the 0h/1h transition for $\phi = 90^{\circ}$.} \label{fig2} \end{figure}

A magnetic-field dependence provides information on the spin states of this first hole. Fig.~\ref{fig2}(a) shows a high-resolution tunnel spectroscopy at the charge transition between the zero- and one-hole state (0h and 1h, respectively).  The measurement was performed at 260 mK under a 6-T in-plane magnetic field, $B_{//}$, parallel to the channel (see fig.~\ref{fig1}(a)) . The corresponding g-factor is labeled as $g_{//}^{0h/1h}$. The lever-arm parameter $\alpha$, relating gate-voltage and electrochemical-potential ($\mu$) variations (i.e.  $\alpha = d\mu/dV_g$), can be obtained from the diamond slopes. We estimate $\alpha = 0.85$ eV/V: such a large value, close to unity, demonstrates the excellent gate control over the channel electrostatic potential. 
Among some ridges invariant under magnetic field, which we ascribe to resonances in the density of states of the leads~\cite{PierreM.2010b}, a strong ridge can  be clearly observed in fig.\ref{fig2}(a), highlighted by a black arrow. This ridge denotes the onset of tunneling via the 1-hole, excited spin state whose separation from the ground state increases linearly with magnetic field as shown in fig.\ref{fig2}(b). This data indicates that the first energy level is two-fold degenerate constituting a spin doublet. The degeneracy lifting between heavy and light holes observed here is a consequence of the strong confinement at the silicon interface.

From the data of fig. \ref{fig2}(b) we extract the in-plane g-factor $g_{//}^{0h/1h}=\frac{\Delta E}{\mu_B B_{//}}$ for magnetic fields parallel to the channel (here $\Delta E$ is the Zeeman energy splitting between the two spin states and $\mu_B$ the Bohr magneton). We find $g_{//}^{0h/1h}\approx 2.3$. The in-plane azimuthal angle, $\phi$, between the magnetic field vector and the channel axis is then varied and the corresponding change in $g_{//}$ recorded. The full angle dependence, shown in figure~\ref{fig2}(c), reveals $g_{//}^{0h/1h}$ values varying between 2.3 and 2.6. The $g$-factor exhibits a strong variation also as a function of the polar angle $\theta$ (see fig.~\ref{fig1}(a)). The $\theta$ dependence measured in the plane orthogonal to the channel axis (i.e. for $\phi=90^{\circ}$), is illustrated in fig.~\ref{fig2}(d). When the magnetic field is aligned along the vertical axis ($\theta=0^{\circ}$), perpendicular to the substrate plane,  the g-factor reaches its smallest value $g_{\perp}^{0h/1h}$=1.5. Moreover, figures~\ref{fig2}(c) and (d) seem to suggest a non-sinusoidal anisotropy. Due to an experimental uncertainty of ($\pm 5^\circ$) on the angle between the magnetic field and the channel axis, however, we cannot conclude on the actual functional form of the angle dependence.

We note that our results are compatible with earlier measurements of the hole $g$-factor in a silicon-nanowire quantum dot~\cite{zwa09}. In that work a g-factor of 2.3 was reported for the first hole and a magnetic field perpendicular to the substrate (and hence to the nanowire). Angular dependences were not addressed in that paper. The g-factors obtained here are strongly anisotropic and mostly below the expectation for heavy holes in a two-dimensional silicon layer, namely $|g|$=6*$|\kappa|$ where the Luttinger parameter $|\kappa|\approx$0.42 in silicon~\cite{win03}.  
This result is likely due to the stronger confinement and the associated mixing between heavy and light hole states. Since confinement is gate-voltage dependent we may expect g-factors to vary with $V_g$, which may open the possibility for spin manipulation g-TMR. This possibility, originally demonstrated for electrons in InGaAs heterostructures~\cite{kato03}, was recently investigated for the case of holes in $SiGe$ self-assembled quantum dots~\cite{ares13}. It was shown that for a fixed, odd number of holes in the quantum dot, a relative small gate-voltage modulation could induce Rabi oscillations at a frequency of $100$ MHz.  

Following a similar approach, we focus on the first-hole regime and investigate the gate-voltage dependence of the parallel and perpendicular components of the hole $g$-factor.
In order to evaluate the maximal variation of the g-factor components within the 1-hole regime we compare the already discussed tunnel spectroscopy measurement of the Zeeman splitting at the $0h/1h$ transition with analogous measurements at the $1h/2h$ transition, i.e. around $V_{g}= -612$ mV. To study the Zeeman splitting at the transition 0h/1h we have recorded a conductance trace at finite positive $V_{b}$ depending on $V_{g}$ marked by a black dashed line in the inset of fig.~\ref{fig2}(a). The same way we have recorded a conductance trace at negative $V_{b}$ (black dashed line in the inset of fig.~\ref{fig2}(a)) to study the 1h/2h transition. The sign of $V_{b}$ has been choosen to have the best contrast between the Zeeman excited state conductance peak and the conductance peaks associated to local density of states in the leads. We emphasize that both sets of measurements , while corresponding to different charge transitions ($0h/1h$ and $1h/2h$), provide information on the Zeeman splitting of the 1-hole ground state. The only difference lies in the electrostatic confining potential, due to different gate voltages.  
The in-plane angular dependence of $g_{//}^{1h/2h}$  is shown in fig.~\ref{fig2}(c) (blue data points). As compared to $g_{//}^{0h/1h}$ (red data points), $g_{//}^{1h/2h}$ is generally smaller and presents a roughly opposite anisotropy. 
While the precise origin of this peculiarity remains unclear, this comparison reveals a strong gate dependence of the $g$-factor in-plane anisotropy. Going from one side to the other of the 1-hole Coulomb-blockade regime, the in-plane g-factor component along the channel is basically constant (around 2.3), while the in-plane component perpendicular to the channel varies from about 1.85 to about 2.6. This strong gate voltage dependence of the $g$-factor could be due to the large variation of the confinement potential enabled by the thin gate oxide $\approx$1 nm. For magnetic-fields applied perpendicular to the plane we find an approximately constant g-factor, i.e. $g_{\perp}^{0h/1h} \approx g_{\perp}^{1h/2h} \approx 1.5$ (shown in the supplementary material S3).

\begin{figure}
 \includegraphics[width=17.5cm]{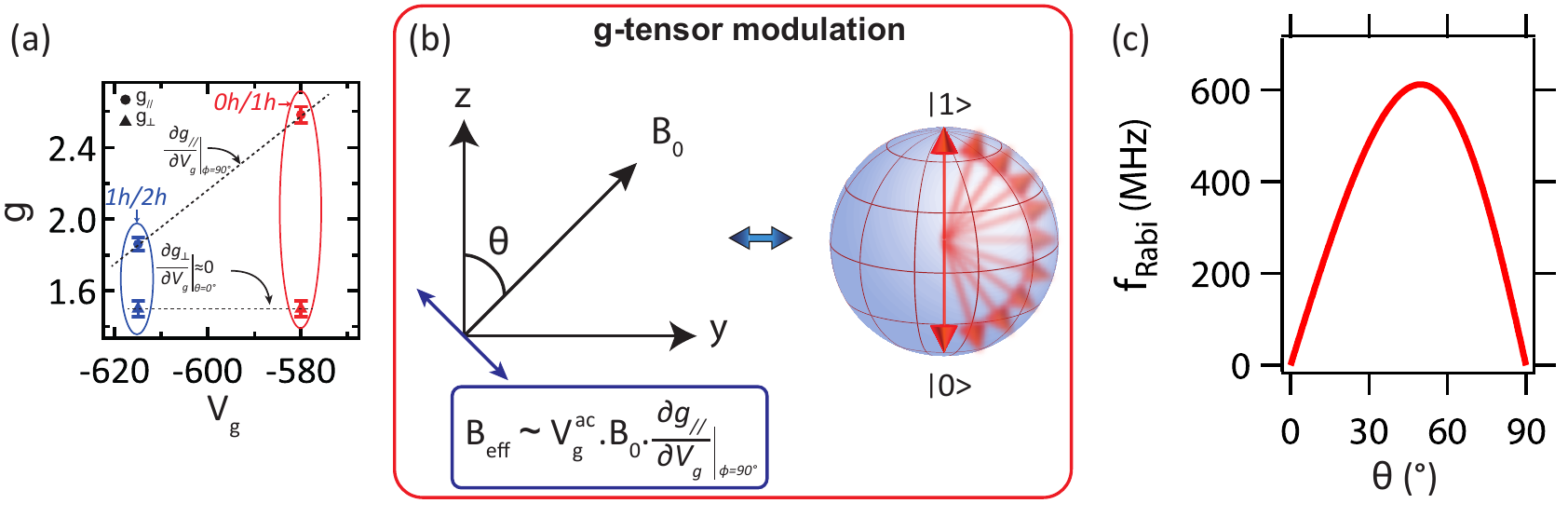} \caption{Illustration of the g tensor modulation technique. A different gate dependence for $g_{\perp}$ and $g_{||}$ combined with an anisotropy a) leads to an equivalent oscillatory magnetic field b) when a time dependent gate voltage is applied together with a static magnetic field $B_0$. This can drive transition between spin states at a speed given by the Rabi frequency. This frequency depends on the system parameters and can exceed 600 MHz in our system. The calculation c) is performed using: $\frac{\partial g_{||}}{\partial V_{g}}\approx 0.025/mV$, $g_{||}=2.6$, $g_{\perp}=1.5$, $B_0$=1 T and $V_{ac}$=10 mV.} 
\label{fig3} 
\end{figure}

The anisotropy and gate dependence of the in-plane g-factors alone are sufficient to allow for g-TMR.  Following ref.~\cite{ares13}, for reasonably strong gate voltage modulation  
$V_{ac}$=10 mV and a static magnetic-field amplitude of 1 T, we calculate the corresponding Rabi frequency as a function of magnetic field orientation.
The results are presented in fig.~\ref{fig3}. First we select $\phi \approx 90^{\circ}$ in order to maximize the gate dependence of $g_{//}$.(From fig.~\ref{fig2}(c), $g_{//}$ goes from $2.6$ at the $0h/1h$ transition  to $1.85$ at the $1h/2h$ transition). Assuming a linear evolution of $g_{//}$ with $V_{g}$, we deduce  $\frac{\partial g_{||}}{\partial V_{g}}\approx 0.025/mV$ while  $g_{\perp}$ is constant at a value of $1.5$, see fig.~\ref{fig3}a). Finally we compute the Rabi frequency at $\phi \approx 90^{\circ}$ with the following expression:
\begin{equation}
f_{Rabi}=\frac{\mu_BV_{ac}}{2h}\left[ \frac{1}{g_{//}}\frac{\partial g_{||}}{\partial V_{g}}\right]\times \frac{g_{//}g_{\perp}B_{//}B_{\perp}}{\sqrt{(g_{//}B_{//})^2+(g_{\perp}B_{\perp})^2}}
\end{equation}

The optimal magnetic field angle $\theta_{max}$ with respect to the device plane is then given as $\theta_{max}=\arctan\left(\sqrt{\frac{g_{//}}{g_{\perp}}}\right) \sim 50^{\circ}$. The corresponding Rabi frequency exceeds $600$ MHz, i.e. significantly larger than Rabi frequencies reported for electric-field driven spin resonance in III-V semiconductor nanostructures (up to $100$ MHz in InSb nanowires~\cite{vandenberg13}). 
In addition to a large Rabi frequency, a conveniently long spin coherence time may be expected in the hole spin qubits studied here. Assuming spin coherence to be limited by hyperfine coupling, we expect $T_2^*$ $\approx 1.5\mu$s for a hole spin in a $[13nm]^3$ silicon quantum dot, which is two orders of magnitude larger than in III-V semiconductors (see supplementary material S5).

In conclusion, the few-hole regime has been achieved in a p-type silicon quantum dot confined within the channel of a field-effect transistor based on industrial level SOI technology. A detailed study of the first accessible hole reveals an anisotropic and gate-dependent hole $g$-factor. The observed gate dependence is maximal for the in-plane g-factor component perpendicular to the channel of the transistor. Moreover we have discussed the revealed anisotropy and gate-dependence in term of possible g-TMR. We estimate that Rabi frequencies as high as several hundred MHz could be reached with realistically strong gate modulation. These frequencies are at least two orders larger than in the case of electron spin qubits using magnetically driven spin resonance~\cite{kop06, pla12,vel14}. In view of the demonstrated potential for fast electric-field driven spin manipulation,  the reduced hyperfine interaction expected from the p-symmetry valence band states, and the full compatibility with mainstream microelectronics technology, the silicon-based p-type quantum dots studied here appear to be an attractive option for spin qubit applications.

\begin{acknowledgement}
The authors thank Y-M Niquet and J. Li for fruitful discussions.The research leading to these results has been supported by the European Community’s seventh Framework under the Grants Agreement No. 323841 (http://www.sispin.eu/) and No. 610637 (http://www.the-siam-project.eu/)  and by the European Research Council under the ERC Grant agreement No. 280043 (HybridNano).
\end{acknowledgement}

\begin{suppinfo}
Additional experimental data about the quantum dot hole shell-filling are provided in the suplementary material S1. An estimation of the hyperfine coupling for holes in silicon is carried in the supplementary material S2.
\end{suppinfo}

\bibliography{BibVoisinetal}

\end{document}